\documentclass[aps,pre,twocolumn,superscriptaddress,floatfix,groupedaddress]{revtex4-1}
\usepackage{amsmath,amsfonts,amssymb}
\usepackage{graphicx,color}
\usepackage{epstopdf}
\usepackage{bm}

\begin{document}
%
%
%
\title{
Two dimensional collective electron magnetotransport, oscillations and chaos in a semiconductor superlattice }
\author{L. L. Bonilla, M. Carretero and A. Segura}
\affiliation{Gregorio Mill\'an Institute, Fluid Dynamics, Nanoscience
and Industrial Mathematics, and Department of Materials Science and Engineering and Chemical Engineering, Universidad Carlos III de Madrid, Legan\'{e}s, Spain}
%
%
%
\begin{abstract}
When quantized, traces of classically chaotic single particle systems include eigenvalue statistics and scars in eigenfuntions. Since 2001, many theoretical and experimental works have argued that classically chaotic single electron dynamics influences and controls collective electron transport. For transport in semiconductor superlattices under tilted magnetic and electric fields, these theories rely on a reduction to a one-dimensional self-consistent drift model. A two-dimensional theory based on self-consistent Boltzmann transport does not support that single electron chaos influences collective transport. This theory agrees with existing experimental evidence of current self-oscillations, predicts spontaneous collective chaos via a period doubling scenario and it could be tested unambiguously by measuring the electric potential inside the superlattice under a tilted magnetic field.
\end{abstract}
\maketitle

\section{Introduction}\label{sec:intro}
\begin{figure}
 \includegraphics[width=8cm]{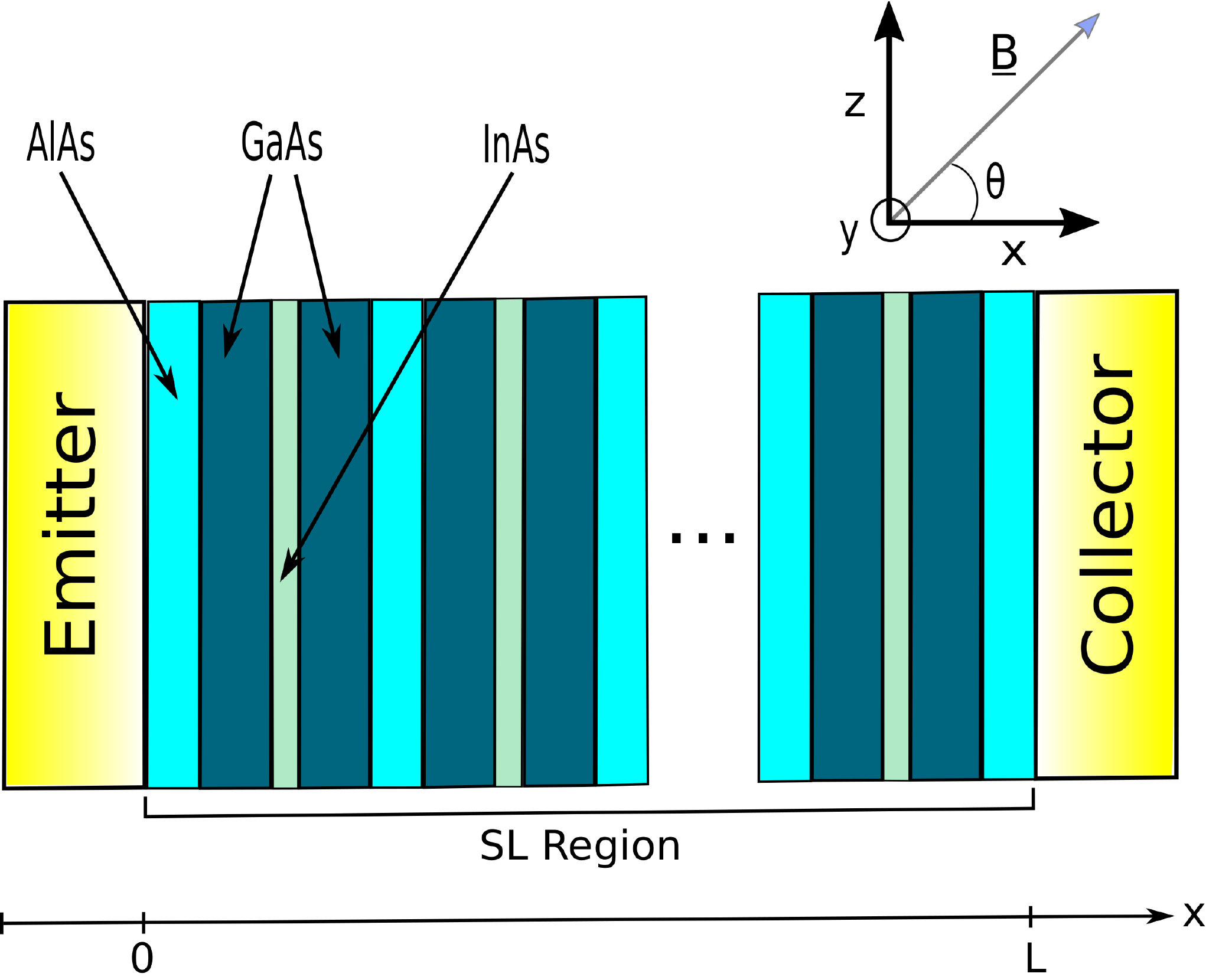}
\caption{
Sketch of the SL device. Each SL period comprises a AlAs barrier, and a well whose central part is made out of InAs and the remainder of GaAs \cite{fro01,fro04}. $L=Nl$. }
\label{fig1}
\end{figure}

Quantum chaos studies the connections between classically chaotic systems and the semiclassical limit of its corresponding quantum mechanical description \cite{gut90,sto99,haa10}. In this fascinating area lying between physics and mathematics, there are conjectures on the different universality classes of energy level spacing distributions \cite{haa10}. Unstable periodic orbits in classically chaotic dynamics appear as scars in wave functions \cite{gut90,sto99}. These are features of classical dynamics with few degrees of freedom. A different problem is to know whether collectivities of classically chaotic systems keep track of single system chaos in quantum transport. Many theoretical and experimental works on electron transport have sought to answer this question in the affirmative. For example, electron dynamics within a semiconductor superlattice (SL) in tilted electric and magnetic external fields (see Figure \ref{fig1}) is classically chaotic, exhibiting stochastic webs and chaotic islands bounded by periodic orbits in their phase space \cite{fro01,fro04}. Will these features of single electron chaos influence collective electron transport in the superlattice?

Fromhold {\em et al} have conjectured that single electron complex dynamics generates resonances between the Bloch and cyclotron frequencies in the collective electron drift \cite{fro01,fro04}. In particular, they generalize the 1970 Esaki-Tsu formula (ETF) for the collective electron drift velocity at zero magnetic field \cite{esa70} to the case of tilted magnetic field. See Appendix \ref{ap1} for a derivation. Then they argue that chaotic diffusion along the stochastic web arising in single electron dynamics produce peaks in the drift velocity. When the resulting drift velocity is inserted in a {\em postulated} self-consistent one-dimensional (1D) model of electron transport, numerical simulations show self-sustained oscillations of the current through the SL that are compared to experiments \cite{fro01,fro04,gre09,ale12}. The origin of the resonant drift velocity peaks has been disputed \cite{sos15}. However, {\em no one seems to have wondered how the multidimensional motion of single electrons in a tilted magnetic field may produce 1D collective electron motion}. Here we derive collective electron motion from a semiclassical self-consistent miniband Boltzmann-Poisson equation, show that collective electron motion is indeed multidimensional, and obtain results that agree with existing experiments. The miniband semiclassical picture is reasonable and Landau levels can be ignored for a wide range of magnetic fields \cite{fow07}. 

Fromhold {\em el al} use the ETF to get the electron current density along the SL growth direction even in the presence of a tilted magnetic field \cite{fro01,fro04}. They assume that the electric field is directed along the SL growth direction. The resulting drift velocity is a function of the electric and magnetic fields and of the tilting angle $\theta$ in Figure \ref{fig1}. Then they describe collective electron transport by a 1D continuity equation for the electron density coupled self-consistently to a Poisson equation for the electric field along the SL growth direction \cite{fro04,gre09}. The resulting system is a diffusionless version of the Kroemer model for the Gunn effect in bulk n-GaAs \cite{kro64,bon10}. The drift velocity may have one or several peaks depending on the tilting angle and the magnetic field \cite{gre09}. This 1D drift-Poisson model is {\em postulated, not consistently derived from a Boltzmann transport equation} (BTE). 

Greenaway {\em et al} solved the 1D drift-Poisson equations by a rough first order discretization of space \cite{gre09}, which converts them in those of a sequential tunneling weakly coupled SL with many more quantum wells. This method regularizes the shock waves appearing in the diffusionless Kroemer model \cite{kni66,bon91} by lattice effects \cite{car01}. It also preserves charge continuity. Gunn-like oscillations due to recycling of charge dipole waves appear in the numerical simulations and multiple peaks in the drift velocity may produce simultaneous coexistence of several dipole waves \cite{gre09}. Note that simultaneous coexistence of several dipole waves has also been observed in numerical simulations of the Kroemer model with a single peaked drift velocity provided the boundary conditions are appropriate \cite{bon97}. Under dc voltage bias in a passive external circuit, self-consistent oscillations are not restricted to a bounded region in parameter space but persist for all voltages larger than critical; see \cite{gre09} and Figure 3(b) of \cite{ale12}. To agree with experimental evidence that self-oscillations exist for bounded voltage intervals (Figure 3(a) of \cite{ale12}), numerical simulations of the diffusionless Kroemer model require coupling of the superlattice to an external resonant circuit representing parasitic impedance, as in Figure 3(c) of \cite{ale12}. Coupling to the external resonant circuit strongly affects the numerically calculated current self-oscillations: their frequency decreases to about 1 GHz and they exist on a finite voltage range that roughly corresponds to the experimentally measured one.

In contrast to all previous works, we use the 2D BTE to obtain a 2D continuity equation for the electron density coupled to a Poisson equation for the electric potential. Numerical simulations under {\em dc} voltage bias conditions show that high magnetic fields confine collective electron motion to a narrow straight channel that goes from the emitter to the receiver contacts. Its inclination is the angle between magnetic field and SL growth direction. In agreement with experiments \cite{ale12}, there are finite voltage intervals within which the current through the SL oscillates in time. Unlike the case of the reduced 1D Kroemer model used in \cite{gre09,ale12}, we do not need coupling to an external resonant circuit to confine the self-consistent current oscillations to a finite voltage interval. However, coupling to the external circuit (which we do not include in the present work) would still be necessary to attain self-oscillations of 1 GHz frequency \cite{ale12}. The self-oscillations of the current arise from recycling of dipole charge waves whose fronts are strongly curved. Period doubling bubbles and period doubling routes to chaos appear. As it could be anticipated from the presence of scattering, collective chaos is dissipative, not conservative as single electron chaos. 

\section{Collective electron transport}\label{sec:2}
Electron collective transport in a miniband of dispersion relation $E(k,k_y,k_z)=\mathcal{E}(k)+\hbar^2(k_y^2+k_z^2)/(2m)$, $\mathcal{E}(k)= \Delta (1-\cos kl)/2$, is described by the BTE
\begin{eqnarray}
\frac{\partial f}{\partial t}\! +\! \frac{\Delta l}{2 \hbar}\sin{kl}\frac{\partial f}{\partial x}\!+\!\frac{\hbar k_z}{m} \frac{\partial f}{\partial z}\!+e\!\left(\frac{F}{\hbar}\!-\!\frac{B}{m}k_y\sin{\theta}\right)\!\frac{\partial f}{\partial k}\nonumber\\
+ e\!\left(\frac{F_z}{\hbar}+\frac{B}{m}k_y\cos{\theta}\right)\!\frac{\partial f}{\partial k_z} =\nu_e(f^B-f) - \nu_p\mathcal{A}f.  \label{eq1}
\end{eqnarray}
Here $\mathcal{A}f= [f(k)-f(-k)]/2$. The distribution function $f(x,z,k,k_z,t)$ is periodic in the wave vector component $k$ along the SL growth direction with period $2\pi/l$, where $l$ is the SL period. $-e<0$, $m$, $-F$, $-F_z$, $B(\cos\theta,0,\sin\theta)$, $\nu_e$, $\nu_p$, are the electron charge, effective mass, electric field components along the $x$ and $z$ axis, the magnetic field, the inelastic and impurity collision frequencies, respectively (see Appendix \ref{ap2}). As a consequence of single electron dynamics, $k_y=eB(x\sin\theta-z\cos\theta)/\hbar$, and electron motion is effectively 2D \cite{fro01}, see Appendix \ref{ap2}. $f^B$ in \eqref{eq1} is \cite{ign87}
\begin{eqnarray}
f^B(k,k_z;n)=\frac{\hbar lL_yn(x,z)}{I_0\!\left(\frac{\Delta}{2k_BT}\right)\!}\sqrt{\frac{\pi}{2mk_BT}}\nonumber\\
\times\exp\!\left(\frac{\Delta}{2k_BT}\cos{kl}-\frac{\hbar^2k^2_z}{2mk_BT}\right)\!,\label{eq2}
\end{eqnarray}
\begin{eqnarray}
n(x,z)=\frac{2}{(2\pi)^2L_y}\int_{-\pi/l}^{\pi/l}\int f^B\, dk\, dk_z\nonumber\\
=\frac{2}{(2\pi)^2L_y}\int_{-\pi/l}^{\pi/l}\int f\, dk\, dk_z,\label{eq3}
\end{eqnarray}
in which $n(x,z)$ is the 3D electron density, and $L_y$ and $L_z$ are the SL extensions along the $y$ and $z$ directions, respectively. The self-consistent electric potential $W$ satisfies the Poisson equation
\begin{eqnarray}
\frac{\partial^2 W}{\partial x^2}+\frac{\partial^2 W}{\partial z^2}=\frac{e}{\varepsilon}(n-N_D),\label{eq4}
\end{eqnarray}
where $N_D$ is the SL doping density and $\varepsilon$ is the SL dielectric constant. Note that $F=\partial W/\partial x$ and $F_z=\partial W/\partial z$. Integration of \eqref{eq1} over the wave vector components produces the charge continuity equation
\begin{eqnarray}
&& e\frac{\partial n}{\partial t}+ \frac{\partial J_{nx}}{\partial x}+\frac{\partial J_{nz}}{\partial z}=0, \label{eq5}\\
&& J_{nx}= \frac{2e}{(2\pi)^2L_y}\int_{-\pi/l}^{\pi/l}\int \frac{\Delta l}{2\hbar}\sin kl\, f\, dk\, dk_z,\nonumber\\ 
&&J_{nz} =\frac{2e}{(2\pi)^2L_y}\int_{-\pi/l}^{\pi/l}\int \frac{\hbar k_z}{m}\,f\, dk\, dk_z. \label{eq6}
\end{eqnarray}

For $B=0$, $F_z=0$, and we can derive the ETF from \eqref{eq1} provided $\nu_p=0$ and $\nu_e=1/\tau$. In this case, we can integrate \eqref{eq1} over $k_z$ and get its 1D version. Appendix \ref{ap1} shows that its solution with initial condition $f(x,k,t_0)=f_0(x,k)$ is
\begin{eqnarray}\nonumber
f(x,k,t)=f_0\!\!\left(x\!-\!\frac{\mathcal{E}(k)}{eF},k\!-\!\frac{eF}{\hbar}(t-t_0)\right)\! e^{-(t-t_0)/\tau}\\
+\int_0^{(t-t_0)/\tau} e^{-\xi}f_{eq}\!\left(k-\frac{eF\tau\xi}{\hbar}\right) d\xi,\label{eq7}
\end{eqnarray}
where $f_{eq}$ is the integral of $f^B$ in \eqref{eq2} over $k_z$. As $t_0\to -\infty$, \eqref{eq7} becomes
\begin{eqnarray}
f_{st}(k)=\int_0^{\infty} e^{-\xi}f_{eq}\!\left(k-\frac{eF\tau\xi}{\hbar}\right)\! d\xi.\label{eq8}
\end{eqnarray}
This stationary 1D electron distribution is equivalent to that used by Esaki and Tsu to derive their drift velocity at zero temperature \cite{esa70}. Inserting \eqref{eq8} in the drift velocity formula produces the generalized ETF:
\begin{eqnarray}\nonumber
&&v_d(F)=\frac{1}{n}\int_{-\pi/l}^{\pi/l} v(k)\, f_{st}(k)\, dk\\
&&\quad=\frac{1}{n}\int_0^\infty e^{-\xi}\int_{-\pi/l}^{\pi/l}v\!\left(k+\frac{eF\tau\xi}{\hbar}\right)f_{eq}(k)\, dk\,d\xi. \label{eq9}
\end{eqnarray}
For our dispersion relation and Boltzmann local equilibrium distribution, we obtain the temperature dependent Esaki-Tsu drift velocity (ETDV) (see Appendix \ref{ap1}):
\begin{eqnarray}
v_d(F)=v_p\frac{2\omega_B\tau}{1+(\omega_B\tau)^2},\quad v_p=\frac{\Delta l I_1\!\left(\frac{\Delta}{2k_BT}\right)}{4\hbar I_0\!\left(\frac{\Delta}{2k_BT}\right)}, \label{eq10}
\end{eqnarray}
in which the Bloch frequency is $\omega_B=eFl/\hbar$. Clearly we cannot obtain the ETF from the BTE \eqref{eq1} in the 2D case when $B\neq 0$, for the electron density depends on the transversal coordinate $z$ and $F_z\neq 0$. This point is further elaborated in Appendix \ref{ap2}.

We now obtain drift-Poisson equations directly from \eqref{eq1}. We assume that Bloch, cyclotron and collision frequencies are of the same order (THz range) and the corresponding terms in \eqref{eq1} dominate all others. Ignoring the latter, we find an approximate distribution function that, inserted in \eqref{eq6}, yields the current density vector (see  Appendix \ref{ap2}),
\begin{eqnarray}
&&J_{nx} = \frac{en\Delta l}{4\hbar}\frac{I_1\!\left(\frac{\Delta}{2k_BT}\right)\!}{I_0\!\left(\frac{\Delta}{2k_BT}\right)\!}\frac{2\nu_e\frac{el}{\hbar}\frac{\partial\Omega}{\partial x} }{\nu_e(\nu_e+\nu_p)+\!\left(\frac{el}{\hbar}\frac{\partial\Omega}{\partial x} \right)^2\! },\label{eq11}\\
&&J_{nz}= \frac{e^2n}{m\nu_e}\frac{\partial\Omega}{\partial z},  \label{eq12}\\
&&\Omega=W-\frac{\hbar^2k^2_{y}}{2me}=W-\frac{eB^2}{2m}(x\sin\theta-z\cos\theta)^2\! .\label{eq13}
\end{eqnarray}
Note that the current density along the $x$ axis, \eqref{eq11}, has the form $env_d$, where $v_d$ is the temperature dependent ETDV for the effective electromagnetic potential $\Omega$ of \eqref{eq13}. Diffusive corrections to the current density vector can be found by the Chapman-Enskog method used to derive drift-diffusion equations in the case $B=0$ \cite{BEP}.

\section{Results}\label{sec:3}
We have solved numerically the 2D drift-Poisson hyperbolic system of equations \eqref{eq4}, \eqref{eq5} and \eqref{eq11}-\eqref{eq13} by the finite volume method \cite{lev02}. The boundary conditions are $J_{nx}=\sigma F$, $J_{nz}=\sigma F_z$, at the contact region $x=0$ (the boundary condition at $x=L=Nl$ has to be added if we include diffusive corrections to the current density), $J_{nz}(x,\pm L_z/2,t)=0$, $W(0,z,t)=0$, and $W(L,z,t)=V$. Initially, the electron density is $n(x,z,0)=N_D+\varepsilon B^2/m$. The current through the SL of Fig.~\ref{fig1} is the sum of electron and displacement currents at the receiving contact:
\begin{eqnarray}
&& I(t)=I_n(t)+I_d(t),\label{eq14}\\
&&I_n(t)=L_y\int_{-L_z/2}^{L_z/2} J_{nx}(L,z,t)\, dz,\label{eq15}\\
&&I_d(t)=L_y\int_{-L_z/2}^{L_z/2}\varepsilon\frac{\partial F}{\partial t}(L,z,t)\, dz.\label{eq16}
\end{eqnarray}
It is interesting to depict the electron current \eqref{eq15} whose time dependent oscillations have larger amplitude than those of $I(t)$.

In our simulations, we use values from experiments \cite{fro04}. $N_D=3\times 10^{22}$ m$^{-3}$, $N=14$, $l=8.3$ nm, $\Delta= 19$ meV, $L_y=L_z=20\,\mu$m and $m=6.1\times 10^{-32}$ kg is the GaAs effective mass. $B$ goes from 0 to 14T. Typical angles are $\theta=0, \pi/6, \pi/3, \pi/2$, whereas typical collision frequencies for high magnetic field are $\nu_e= 0.7$ THz, $\nu_p= 7$ THz, so that $\sqrt{\nu_e(\nu_e+\nu_p)}= 2.3$ THz. The inelastic frequency $\nu_e$ is known to decrease for increasing magnetic fields \cite{fow06}, so we have set a larger frequency $\nu_e=1.35$ THz for smaller values $B<2T$. To get a peak current of 25 mA at the onset of oscillations as in the experiments \cite{ale12}, we set 90 K as the effective temperature instead of the lattice temperature of 4.2 K \cite{gre09}. 

\begin{figure}
\includegraphics[width=8.5cm]{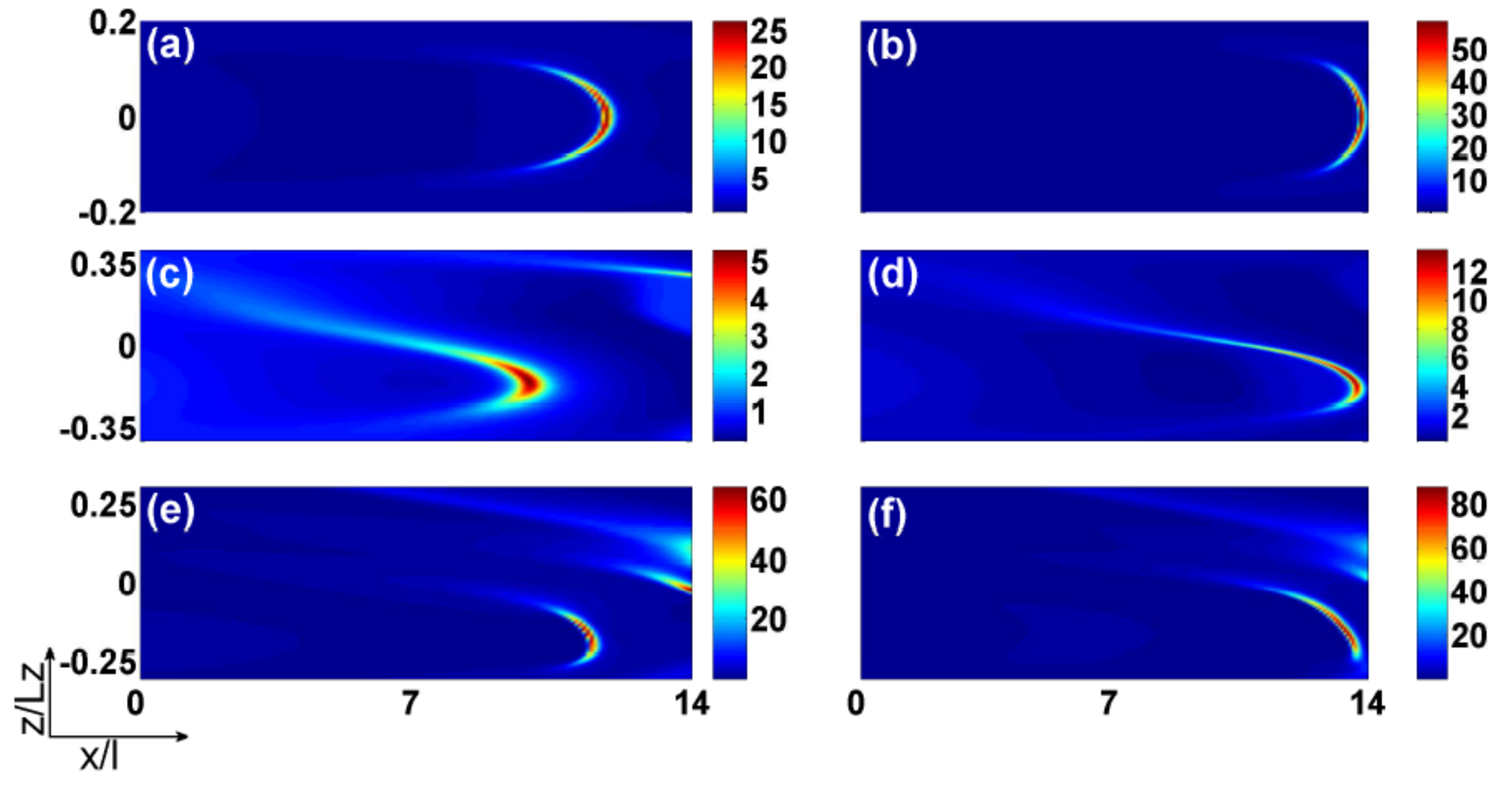}
\caption{
Snapshots of the density profiles at two different times during an oscillation period for (a)-(d): $V=0.12$ V, $B=0.1$ T and tilting angles $\theta=0$, 60$^o$ and (e)-(f) $V=0.32$ V, $B=1$T and $\theta=85^o$. Note that in these last two panels several dipole waves seem to coexist simultaneously. Here $N=14$, $\sigma_c=0.13/(\Omega$ cm).}
\label{fig2}
\end{figure}

\begin{figure}
\includegraphics[width=8cm]{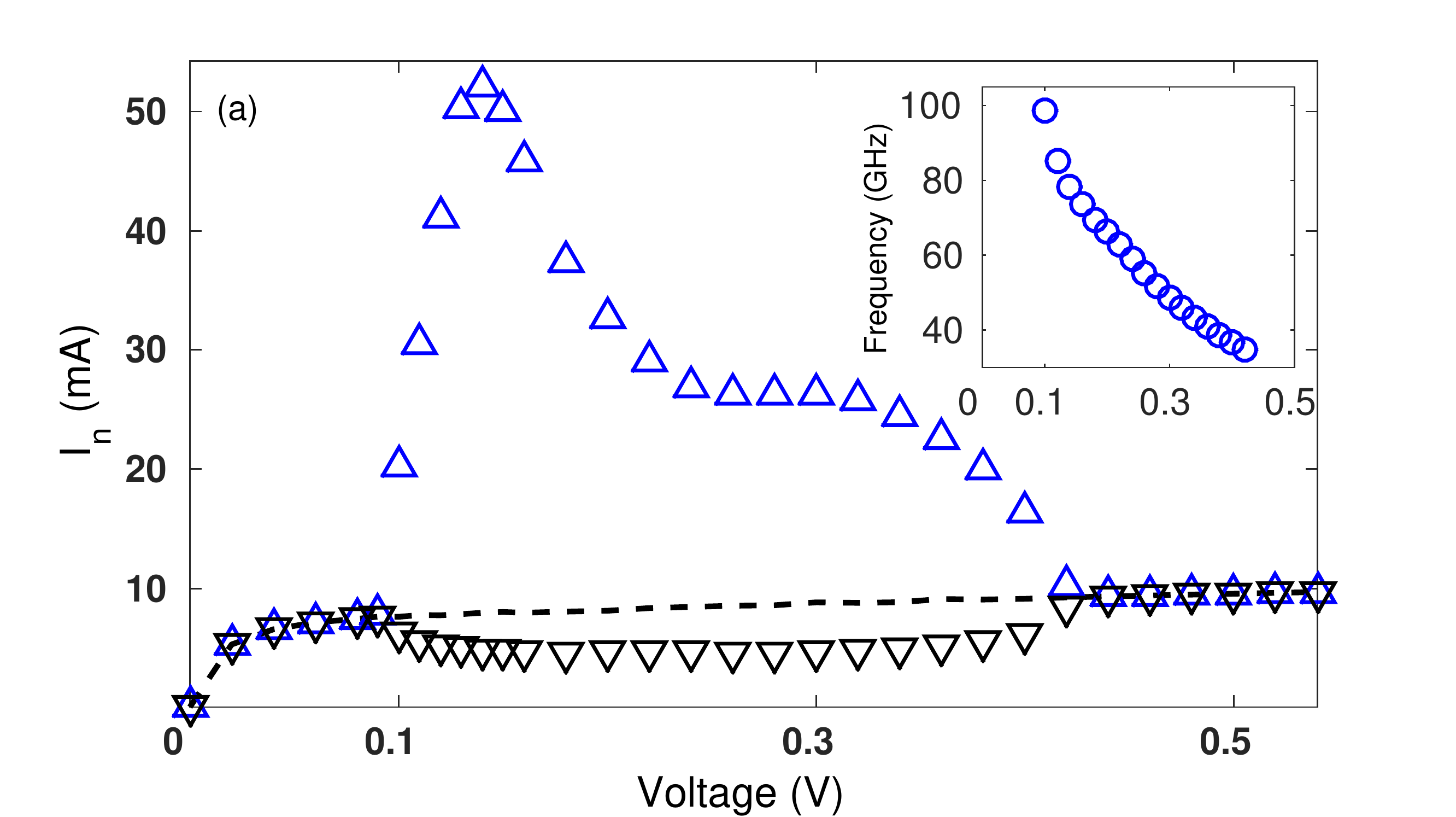}
\includegraphics[width=8cm]{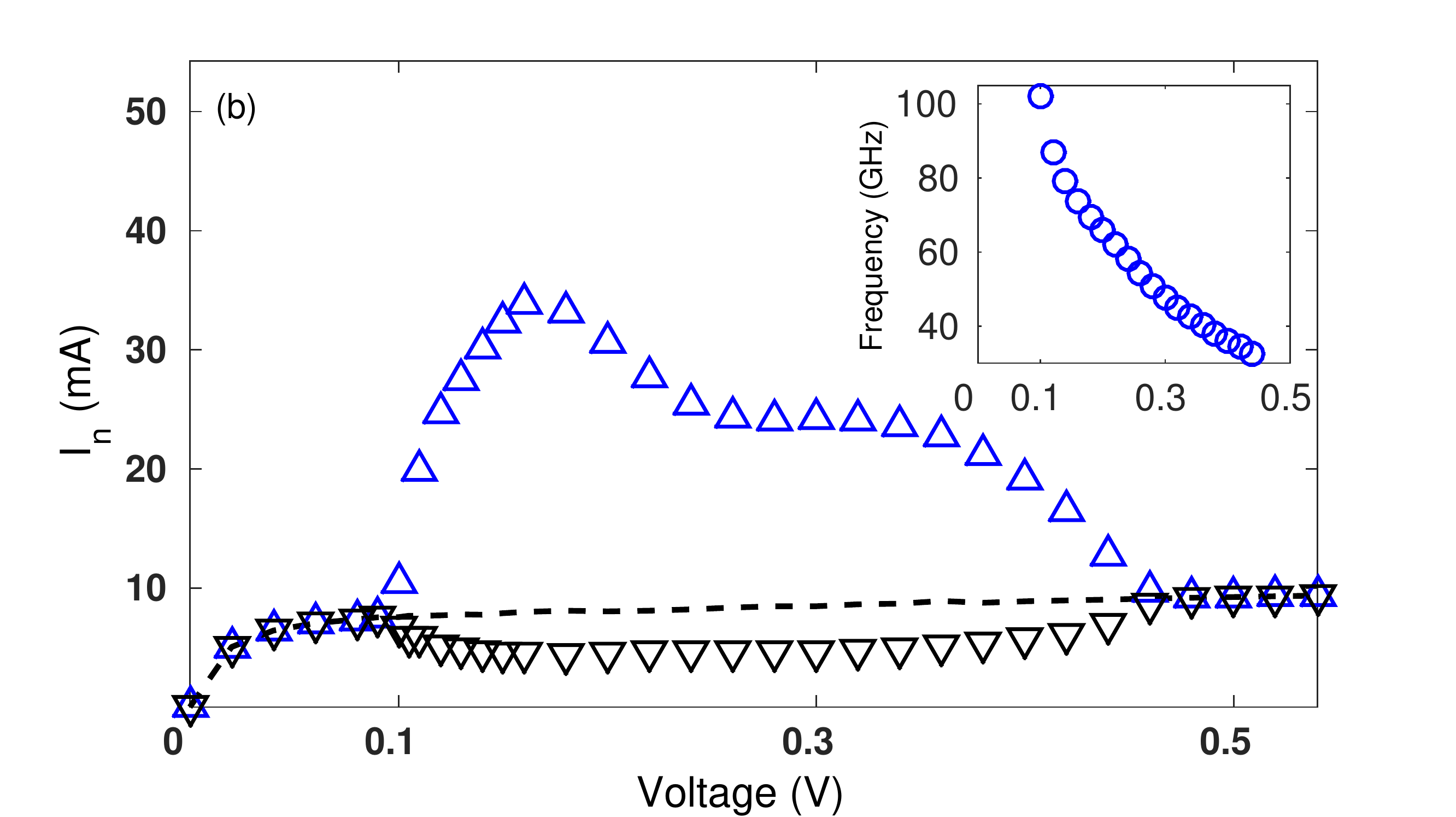}
\includegraphics[width=8cm]{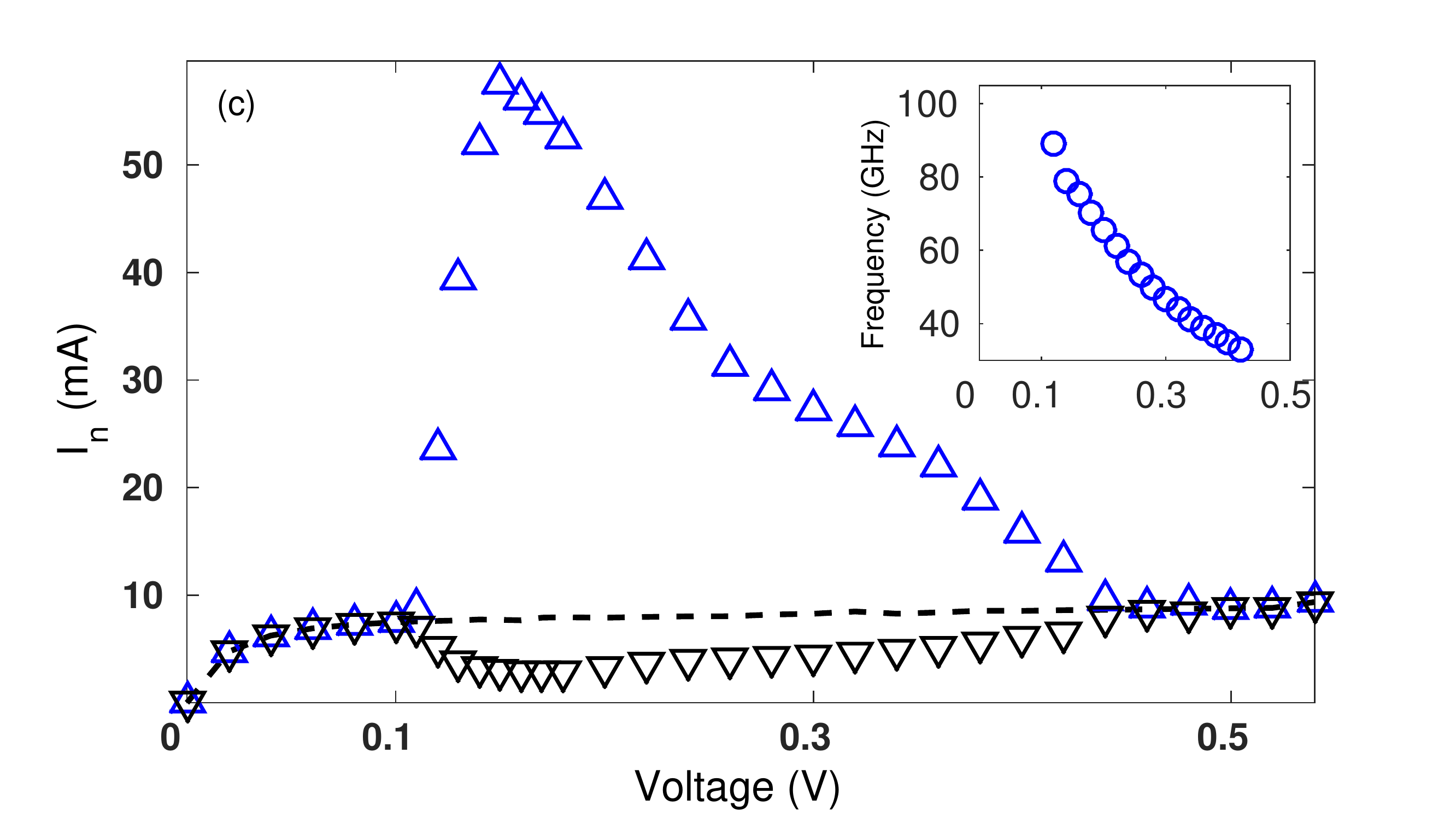}
\includegraphics[width=8cm]{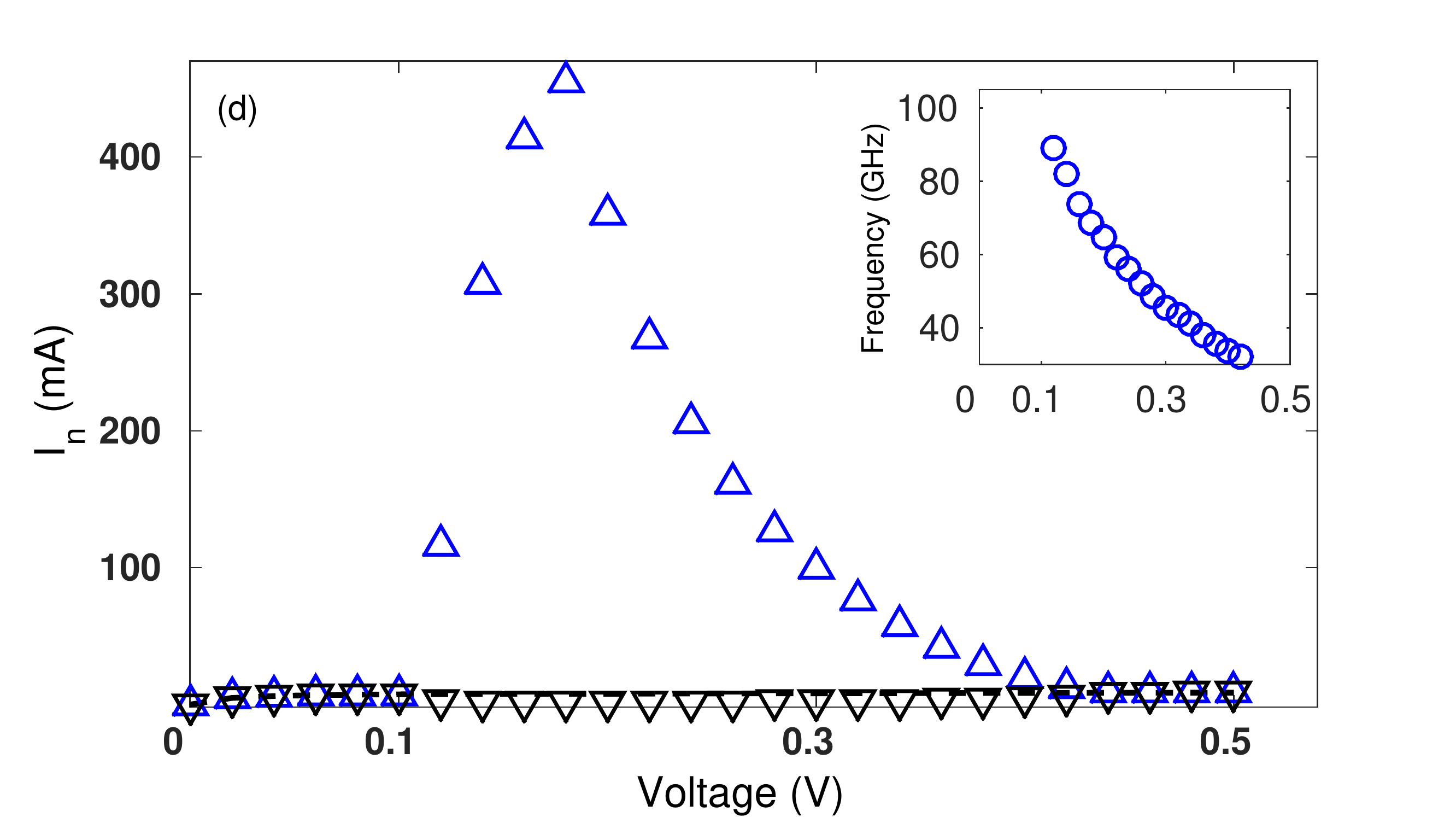}
\caption{$I_n-V$ characteristics showing maxima, minima and mean values of the electron current during oscillations, and frequency vs voltage curves for $B=0.1$ T and $\theta$ equal to (a) 0, (b), $\pi/6$, (c) $\pi/3$, and (d) $\pi/2$. }
\label{fig3}
\end{figure}

For small $B$ and voltage above critical, there are time periodic oscillations of the current due to repeated recycling and motion of curved charge dipole domains, as shown by Figure \ref{fig2}. Note that the electron density in the dipole wave grows significantly as it approaches the collector contact and starts disappearing there. This is also a feature of the 1D Gunn effect in bulk n-GaAs \cite{bon10} and of Gunn-like oscillations in weakly or strongly coupled SLs \cite{BGr05}. As in the 1D Gunn effect \cite{bon10,kni66,bon91}, the dipole wave exhibits a large electron density in its sharp backfront (negative charge) and a very small electron density (positive charge) in its wide forefront; see Figure \ref{fig2}(c). Increasing the tilting angle simply breaks the reflection symmetry of the fronts in Figures \ref{fig2}(a) and (b). 

\begin{figure}
\includegraphics[width=8.5cm]{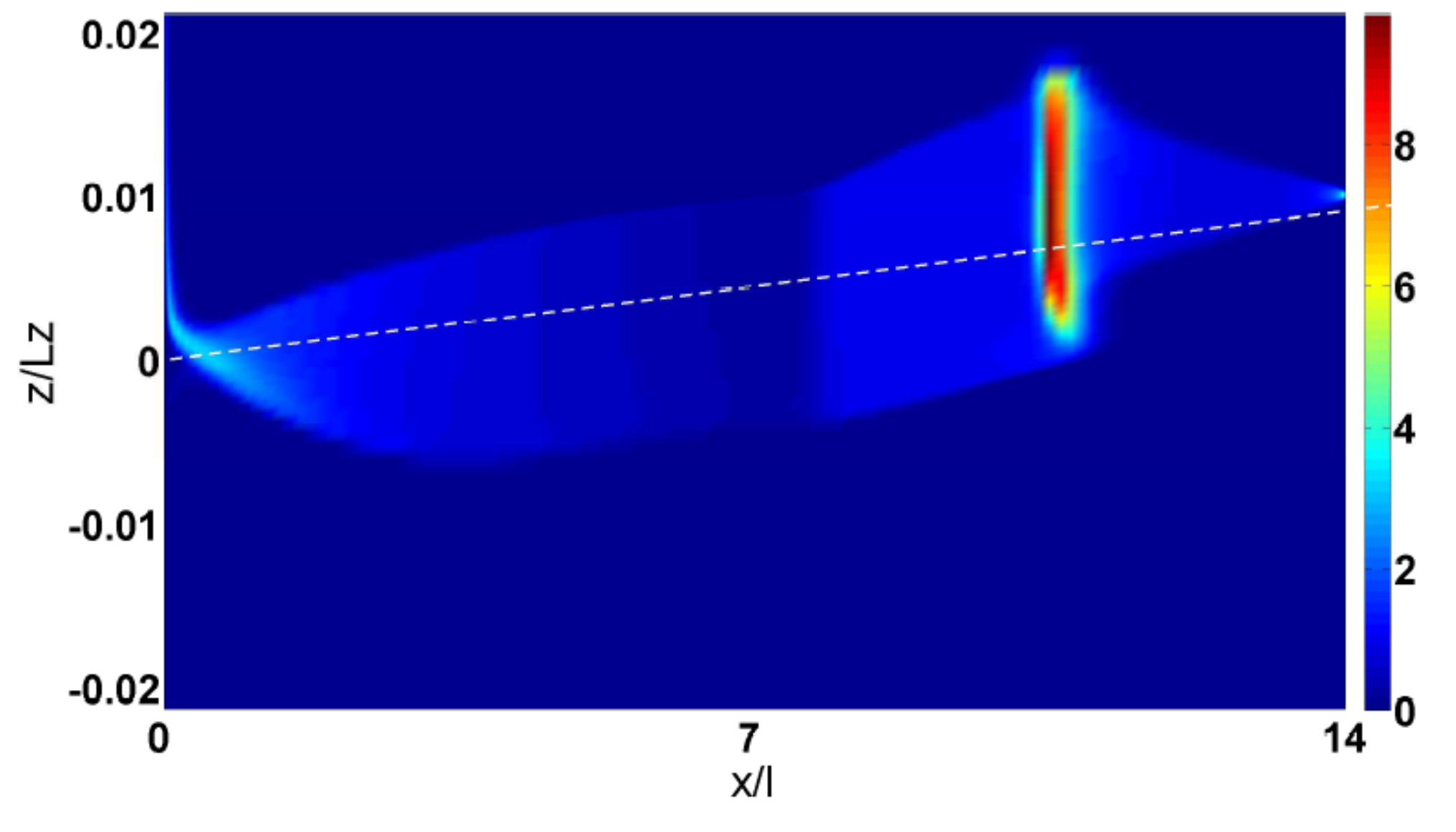}
\caption{
Zoom of the electron density profile during oscillations for $B=2$ T, $\theta=\pi/3$ and $V=0.09$ V. The electron density is almost zero outside a narrow channel with an inclination of $\pi/3$ (marked by the dotted white line $z=\sqrt{3}x$). }
\label{fig4}
\end{figure}

\begin{figure}
\includegraphics[width=8cm]{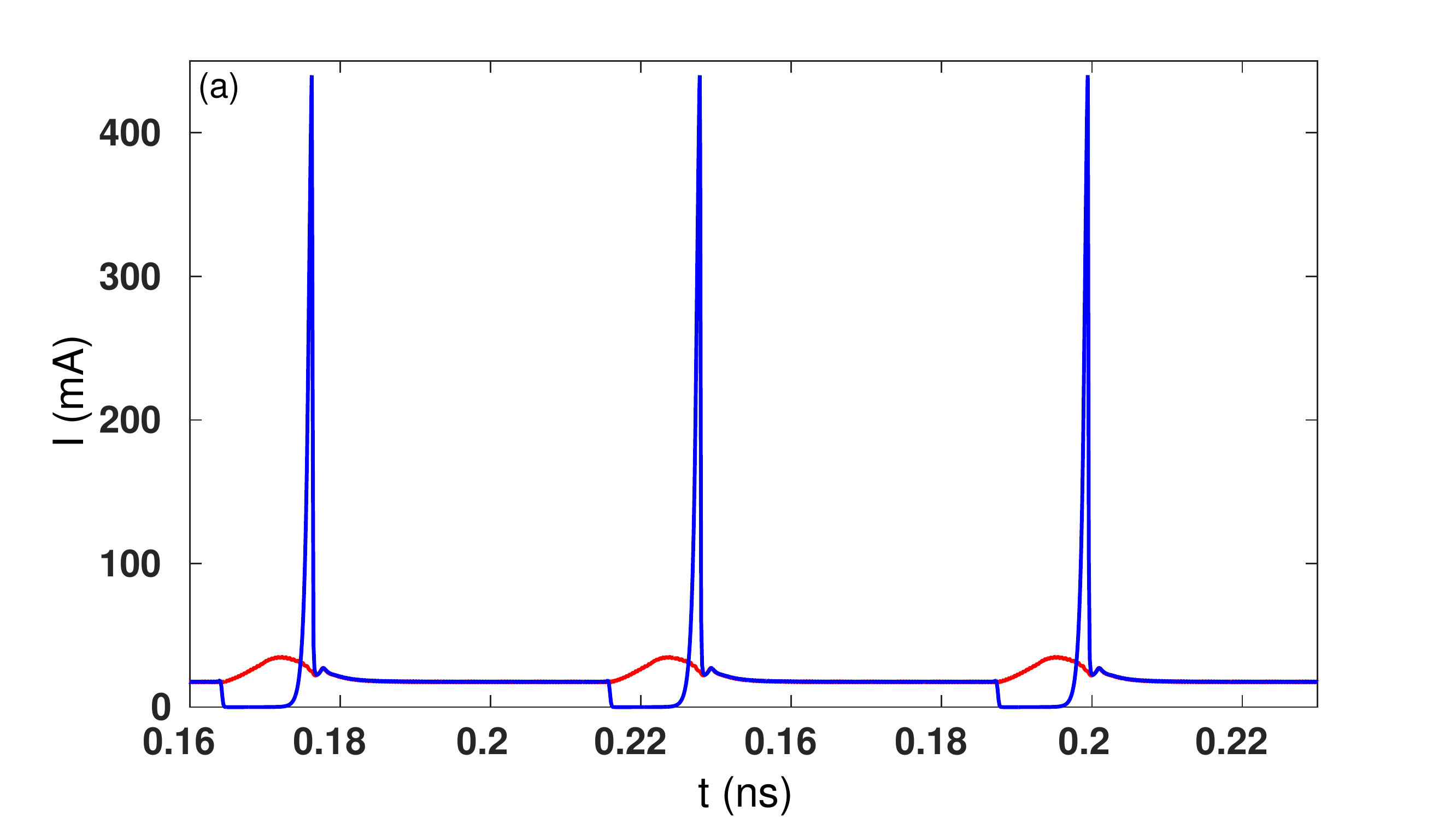}
\includegraphics[width=8cm]{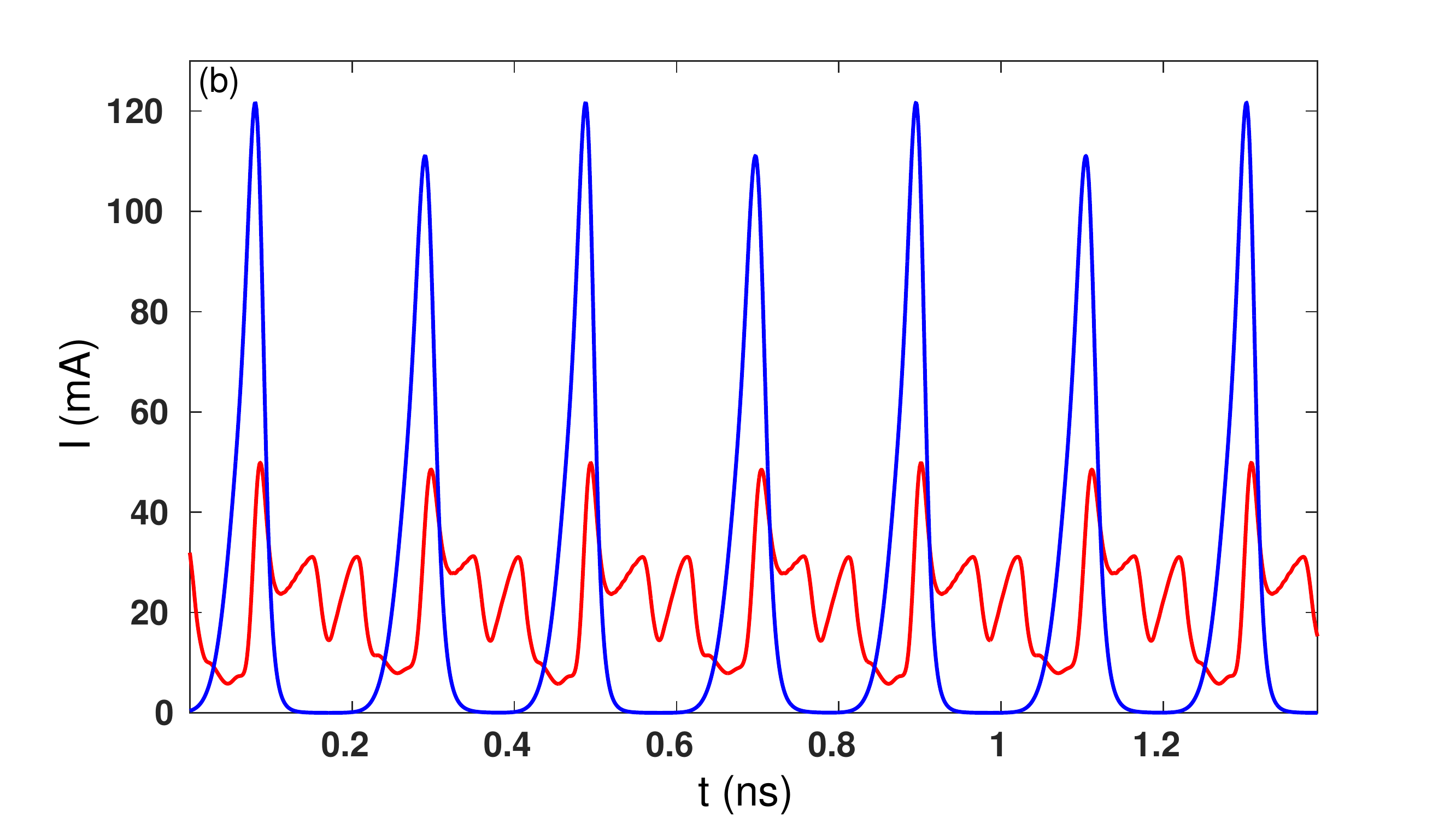}
\caption{Total (red) and electron (blue) current versus time for $B=7$T, $\sigma_c=1.83/(\Omega$ cm) and $\theta=\pi/2$ for (a) $V=0.9$ V, and (b) $V= 1.8476$ V. }
\label{fig5}
\end{figure}

Figure \ref{fig3} shows the $I_n-V$ curve and frequency vs voltage curves for $B=0.1$ T and $\theta=0,\,\pi/6,\,\pi/3,\,\pi/2$. We have displayed maxima, minima and average of the electron current self-oscillations. They begin and end at supercritical Hopf bifurcations issuing from the stationary state. As $B$ increases, electron motion becomes confined in a narrow channel of slope $\tan\theta$, $0 \leq \theta<\pi/2$, corresponding to $k_y=0$, see Figure \ref{fig4}. Despite the increasing magnetic field, the effective potential $\Omega$ of \eqref{eq13} remains close to the electric potential $W$. Electrons move collectively in a quasi 1D manner acted upon by an effective field $(F(x,x\tan\theta,t),0,F_z(x,x\tan\theta,t))$ {\em that is not directed along the growth direction}. For much narrower SLs, the front of the dipole wave may reach the side walls before arriving at the anode. This case requires a separate study to ascertain the effect of the side boundary condition on the dipole waves.

At $\theta=\pi/2$, electron motions along the $x$ and $z$ directions are uncoupled and $F_z=0$ if the initial electron density is independent of $z$. Then we get an effective 1D drift-Poisson system of equations along the $x$ axis. The $I-V$ characteristics  for any $\theta$ are similar to those for low magnetic field, but the collective electron dynamics may become more complex for $B\geq 2$ T and the end of the oscillation may come at finite amplitude. As shown in Figure \ref{fig5}, the electron current is much higher (and provides better contrast) than the total current because the displacement current tends to oppose the former. This is most noticeable for voltages just above critical as in Figure \ref{fig5}(a). For a larger voltage, the electron current in Figure \ref{fig5}(b) shows clear period doubling but the same phenomenon is harder to appreciate for the total current.

\begin{figure}
\includegraphics[width=8cm]{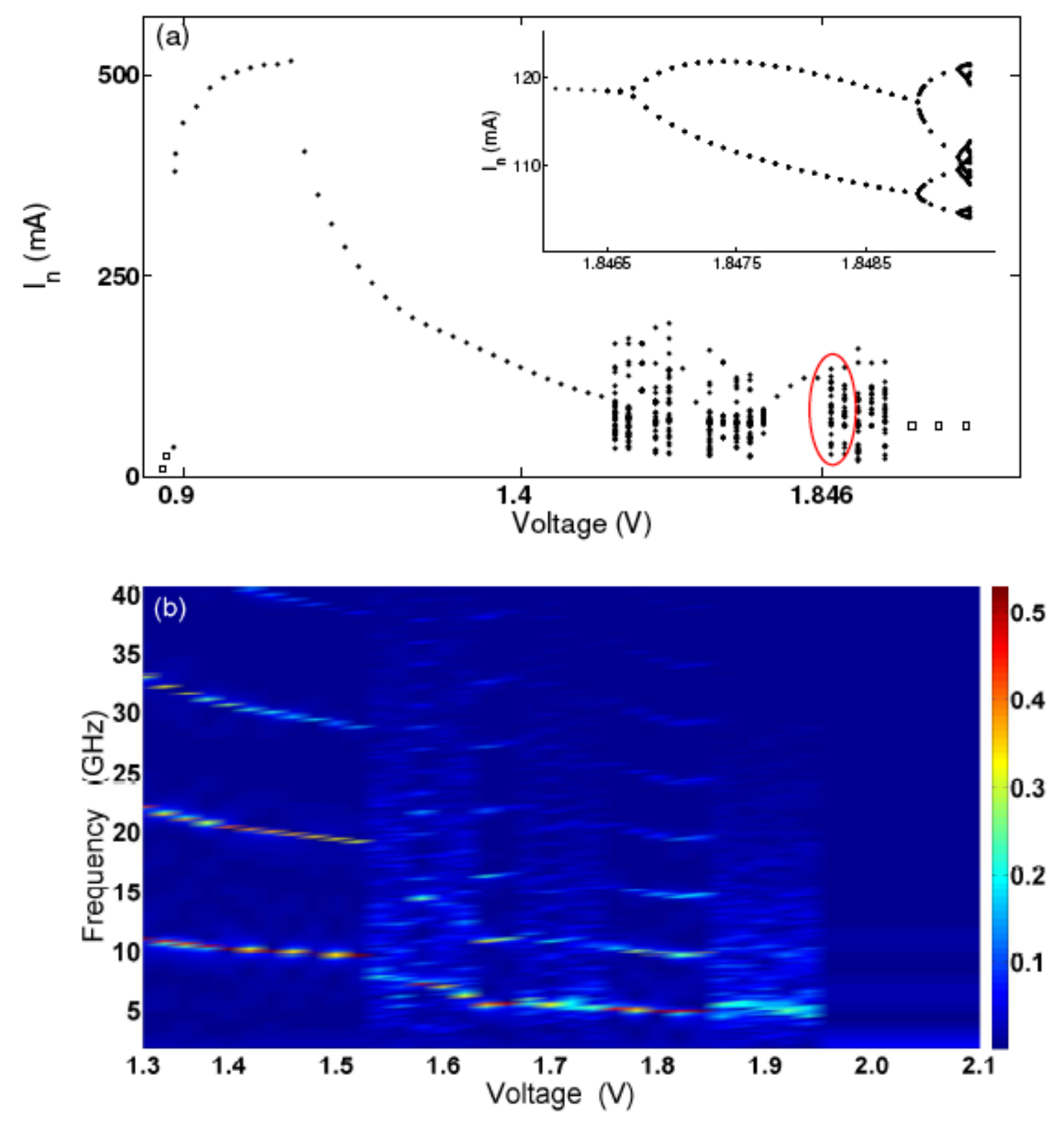}
\caption{(a) Poincar\'e map of the electron current vs voltage depicting period doubling bubbles and period doubling cascades for the voltage interval corresponding to current oscillations density profile during oscillations. The squares correspond to the current at the stable stationary state at lower and higher voltages at which there are no oscillations. The inset is a zoom of the marked region showing the beginning of a Feigenbaum period doubling cascade to chaos. (b) Density plot of frequency vs voltage. Parameters as in Figure \ref{fig5}.} 
\label{fig6}
\end{figure}

The Poincar\'e mapping of Figure \ref{fig6}(a) and the frequency plot of Fig.~\ref{fig6}(b) confirm the complex collective electron dynamics found for a larger magnetic field, $B=7$T, and $\theta=\pi/2$. There are period doubling bubbles in voltage subintervals and period doubling cascades ending in dissipative chaotic attractors. In all cases, and as it happens in Figure \ref{fig2}, the voltage interval of current self-oscillations is finite, which agree qualitatively with experimental observations (see Fig.~3(a) of \cite{ale12}). In contrast, calculations based on the ETF and a 1D discrete drift model produce voltage intervals of current oscillations that do not end (see Fig.~3(b) of \cite{ale12} and \cite{gre09}) unless an external resonant circuit is added to the system (see Fig.~3(c) of \cite{ale12}). The resonant circuit represents parasitic impedance and is also responsible for lowering the oscillation frequency from about 10 GHz to 1 GHz, as observed in experiments \cite{ale12}. Numerical simulations of the ETF based 1D drift-Poisson (Kroemer) model indicate that quasiperiodic and chaotic oscillations appear only for superimposed {\em dc} and {\em ac} voltage biases \cite{kor13}. Apparently, and unlike our results based on 2D calculations, spontaneous chaos under dc voltage bias was not observed in 1D simulations \cite{kor13}, although one period doubling bubble was found in \cite{sel16}. Note that period doubling bubbles and the Feigenbaum route to chaos have been found in  simulations of weakly coupled superlattices \cite{ama02,rui17}. Figures \ref{fig2} and \ref{fig4} further illustrate the multidimensional collective motion of electrons that clearly cannot be captured by 1D averages of the group velocity projected onto the SL growth direction \cite{fro01,fro04,sos15,gre09,kor13}.

Existing experiments measure time resolved current and current--voltage characteristics. However, the current through the device is a scalar magnitude that averages out spatial information and simulations of many different models may produce similar current traces. Moreover, qualitative features appearing in simulations, such as simultaneous coexistence of dipole waves, can have different causes: several peaks in the drift velocity \cite{gre09} or boundary conditions for a model with a single peak \cite{bon97,bon10}. This is already clear from simulations of the 1D \cite{bon97} and the 2D \cite{bon01,bon03} Kroemer model. Thus we would need to measure or reconstruct 2D maps of electron density or electric potential to test unambiguously our predictions. It would also be helpful to have more complete measurements that minimize parasitic impedance effects \cite{ale12} and provide current-voltage curves with a finer voltage grid (similar to our figure \ref{fig3}) for more values of the tilting angle. These more precise measurements would allow discriminating which features of the oscillations are intrinsic to a device free from parasitic impedance effects.

\section{Conclusions}\label{sec:4} 
Collective electron motion in a semiconductor superlattice under combined magnetic and self-consistent electric fields is intrinsically multidimensional. High magnetic fields confine electrons to a narrow channel tilted with the same angle as the magnetic field forms with the superlattice growth direction. Under dc voltage bias, scattering transforms the complex conservative motion of single electrons into spontaneous oscillations of the current that may be periodic, quasiperiodic or chaotic in nature. These oscillations exist on finite voltage intervals and may appear for modest magnetic fields (see Figure \ref{fig2}). Our predictions capture qualitative features observed in experiments, including that, under {\em dc} voltage bias, current self-oscillations are confined to finite voltage intervals. 

In contrast with ours, previous theory extends the Esaki-Tsu formula to the 2D configuration resulting from a tilted magnetic field and finds a multipeaked drift velocity \cite{fro01,fro04,sos15,gre09,kor13,sel16}. Assuming that the electric field is directed along the superlattice growth direction (which is not the case, as shown in Appendix \ref{ap2}), the obtained drift velocity is then inserted in a 1D drift-Poisson Kroemer model. The latter is not derived from Boltzmann-Poisson equations or any more general theory. When the 1D Kroemer model is coupled to an external resonant circuit, numerical solutions of the resulting model produce finite intervals of self-oscillations whose frequency agrees with experimental observations \cite{ale12}. 

There exist measurements of time resolved current traces and current--voltage characteristics. However, the current through the device averages out space information and different theories may produce similar values. The obvious way to test our predictions unambiguously is to reconstruct the 2D electric potential and/or electron density inside the superlattice directly from experiments. Hopefully our results may stimulate new experiments that provide more abundant data on the shape of the current self-oscillations (with less parasitic impedance effects) for more tilting angles and voltages as well as maps of 2D electric potential inside the superlattice.

 \acknowledgements  
This work has been supported by the Ministerio de Econom\'\i a y Competitividad grant  MTM2014-56948-C2-2-P.

\appendix
\setcounter{equation}{0}
\renewcommand{\theequation}{A.\arabic{equation}}
\section{Derivation of the Esaki-Tsu formula}
\label{ap1} 
Here we derive the Esaki-Tsu formula (ETF) from kinetic theory. We start with the 1D Boltzmann equation with relaxation-time collisions:
\begin{eqnarray}
\frac{\partial f}{\partial t} +v(k)\,\frac{\partial f}{\partial x}
+ \frac{eF}{\hbar}\,\frac{\partial f}{\partial k}=\frac{f_{eq}(k)-f}{\tau},\label{s1}
\end{eqnarray}
in which $-e<0$ and $-F$ are the electron charge and the electric field, respectively. $f_{eq}(k)$ is the appropriate local equilibrium function and $v(k)=\hbar^{-1}d\mathcal{E}/dk$ is the group velocity corresponding to the dispersion relation $\mathcal{E}(k)$. The characteristic equations of \eqref{s1} are
\begin{eqnarray}
\frac{dx}{dt}&=&v(k),\label{s2}\\ 
\frac{dk}{dt}&=&\frac{eF}{\hbar},\label{s3}\\
\frac{df}{dt}&=&\frac{f_{eq}(k)-f}{\tau}.\label{s4}
\end{eqnarray}
The solution of \eqref{s1} with initial condition $f(x,k,t_0)=f_0(x,k)$ follows from the solution of \eqref{s2}-\eqref{s4}:
\begin{eqnarray}
&&x(t;x_0,k_0)=x_0+\frac{1}{eF}\mathcal{E}\!\left(k_0+\frac{eF}{\hbar}(t-t_0)\right)\!, \label{s5}\\
&&k(t;x_0,k_0)=k_0+\frac{eF}{\hbar}(t-t_0),\label{s6}\\
&&f(t;x_0,k_0)=f_0(x_0,k_0)\, e^{-(t-t_0)/\tau}\nonumber\\
&&\quad+\frac{1}{\tau}\int_{t_0}^t e^{-(t-s)/\tau}f_{eq}\!\left(k_0+\frac{eF}{\hbar}(s-t_0)\right) ds.\label{s7}
\end{eqnarray}
After changing variables, $\xi=(t-s)/\tau$, the last equation can be rewritten as
\begin{eqnarray}
f(t;x_0,k_0)=f_0(x_0,k_0)\, e^{-(t-t_0)/\tau}\quad\quad\quad\quad\quad\quad\nonumber\\
+\int_0^{(t-t_0)/\tau}\! e^{-\xi} f_{eq}\!\left(k_0+\frac{eF}{\hbar}(t-t_0-\tau\xi)\right)\! d\xi.\label{s8}
\end{eqnarray}
To get the solution of the initial value problem for \eqref{s1}, we have to solve first \eqref{s5} and \eqref{s6} for $x_0$ and $k_0$ as functions of $x$ and $k$:
\begin{equation}
k_0=k-\frac{eF}{\hbar}(t-t_0),\quad x_0=x-\frac{\mathcal{E}(k)}{eF}.\label{s9}
\end{equation}
Inserting this result in \eqref{s8}, we get
\begin{eqnarray}
&&f(x,k,t)=f_0\!\left(\!x-\frac{\mathcal{E}(k)}{eF},k-\frac{eF}{\hbar}(t-t_0)\!\right)\! e^{-(t-t_0)/\tau}\nonumber\\
&&\quad+\int_0^{(t-t_0)/\tau} e^{-\xi}f_{eq}\!\left(k-\frac{eF\tau\xi}{\hbar}\right) d\xi.\label{s10}
\end{eqnarray}
As $t_0\to-\infty$, \eqref{s10} produces the stationary solution of \eqref{s1}:
\begin{eqnarray}
f_{st}(k)=\int_0^{\infty} e^{-\xi}f_{eq}\!\left(k-\frac{eF\tau\xi}{\hbar}\right) d\xi.\label{s11}
\end{eqnarray}
The drift velocity is 
\begin{eqnarray}
&&\!\!v_d(F)=\frac{1}{n}\int_{-\pi/l}^{\pi/l} v(k)\, f_{st}(k)\, dk\nonumber\\
&&\!\!=\!\frac{1}{n}\int_0^\infty e^{-\xi}\!\int_{-\pi/l}^{\pi/l}\!v\!\left(k+\frac{eF\tau\xi}{\hbar}\right)\!f_{eq}(k) dk d\xi. \label{s12}
\end{eqnarray}
This is the generalized ETF. For a Boltzmann distribution, 
\begin{eqnarray}
f_{eq}(k)=\frac{nl}{2\pi I_0\!\left(\frac{\Delta}{2k_BT}\right)}\exp\!\left[\frac{\Delta}{2k_BT}\cos kl\right]\!,\label{s13}\\
 n=\int_{-\pi/l}^{\pi/l} f_{eq}(k)\, dk,\nonumber
\end{eqnarray}
corresponding to the tight-binding dispersion relation
\begin{equation}
\mathcal{E}(k)=\frac{\Delta}{2}(1-\cos kl), \label{s14}
\end{equation}
\eqref{s12} yields the drift velocity
\begin{eqnarray}
v_d(F)\!&=&\!\frac{\Delta l^2}{4\pi\hbar I_0\!\left(\frac{\Delta}{2k_BT}\right)}\int_0^\infty\!\! e^{-\xi}\int_{-\pi/l}^{\pi/l}\sin(kl+\omega_B\tau\xi)\nonumber\\
&\times&\exp\!\left[\frac{\Delta}{2k_BT}\cos kl\right] dk\,d\xi. \label{s15}
\end{eqnarray}
Here $k_B$ is the Boltzmann constant. Splitting the sine function in \eqref{s15}, we obtain the temperature dependent Esaki-Tsu drift velocity (ETDV)
\begin{eqnarray}
v_d(F)\!&=&\!\frac{\Delta l I_1\!\left(\frac{\Delta}{2k_BT}\right)}{2\hbar I_0\!\left(\frac{\Delta}{2k_BT}\right)}\int_0^\infty e^{-\xi}\sin(\omega_B\tau\xi)\,d\xi\Longrightarrow\nonumber\\
v_d(F)\!&=&\! v_p\frac{2\omega_B\tau}{1+(\omega_B\tau)^2},\label{s16}\\ 
v_p\!&=&\!\frac{\Delta l}{4\hbar}\frac{I_1\!\left(\frac{\Delta}{2k_BT}\right)}{I_0\!\left(\frac{\Delta}{2k_BT}\right)}, \label{s30}
\end{eqnarray}
in which $\omega_B=eFl/\hbar$ is the Bloch frequency. As $T\to 0$, we use $I_0(z)\sim e^z/\sqrt{2\pi z}$ and \eqref{s13} becomes
\begin{eqnarray*}
f_{eq}(k)\sim nl\sqrt{\frac{\Delta}{4\pi k_BT}} \exp\!\left[\frac{\Delta}{2k_BT}(\cos kl-1)\right]\!. 
\end{eqnarray*}
Only those $k$ that make $\cos kl\sim 1$ contribute to moments calculated with this expression. Thus, we may further approximate
\begin{eqnarray}
\!\!\! f_{eq}(k)\sim n\sum_{j=-\infty}^\infty\sqrt{\frac{\Delta l^2}{4\pi k_BT}} \exp\!\left[\frac{-\Delta (kl-2j\pi)^2}{4k_BT}\right]\!\!.\quad\quad \label{s17}
\end{eqnarray}
As $k_BT\ll\Delta$, we obtain
\begin{eqnarray}
f_{eq}(k)\sim n\sum_{j=-\infty}^\infty\delta(k-2j\pi). \label{s18}
\end{eqnarray}
Then the zero-temperature stationary solution \eqref{s11} is
\begin{eqnarray}
\!\!\! f_{st}(k)\!=n\int_0^{\infty}\! e^{-\xi}\!\!\sum_{j=-\infty}^\infty\!\!\delta\!\left(k-2j\pi-\frac{eF\tau\xi}{\hbar}\right)\! d\xi,\label{s19}
\end{eqnarray}
which yields the drift velocity
\begin{eqnarray}
v_d(F)\!&=&\!\int_0^{\infty}\!e^{-\xi}\!\int_0^{\infty}\! e^{-\xi}v(k)\!\!\sum_{j=-\infty}^\infty\!\delta\!\left(k-2j\pi-\frac{eF\tau\xi}{\hbar}\right)\! dk d\xi\nonumber\\
&=&\int_0^{\infty} e^{-\xi}\, v\!\left(\frac{eF\tau\xi}{\hbar}\right) d\xi. \label{s20}
\end{eqnarray}
This is the ETF, which yields \eqref{s16} with $v_p=\Delta l/(4\hbar)$ instead of \eqref{s30} for the tight binding dispersion relation \eqref{s14}. Note that $k=eF\tau\xi/\hbar$ in the ETF \eqref{s20} is the solution of the characteristics equation \eqref{s3} with initial condition $k_0=0$ for $t_0=0$. If we solve the characteristics equation \eqref{s3} with initial condition $k_0$ selected out of a Gaussian distribution with variance $4k_BT/(\Delta l^2)$, then we reconstruct the distribution \eqref{s17}. In the absence of a magnetic field, this justifies Fromhold {\em et al} \cite{fro04} usage of the ETF for nonzero ultralow temperatures. It is obvious that the derivation we have presented in this appendix does not hold for truly 2D electron density and electric potential.

\setcounter{equation}{0}
\renewcommand{\theequation}{B.\arabic{equation}}
\section{Derivation of the continuity equation to leading order}\label{ap2}
\subsection{Single electron equations}
We can derive a drift-diffusion equation from the BTE \eqref{eq1} by using the Chapman-Enskog method as in Reference \cite{BEP}. In this section, we will derive the leading order of such an equation from a Boltzmann-Poisson system that includes a magnetic field for electrons in a miniband with tight-binding dispersion relation.  The equations of motion of a single electron in combined electric and magnetic fields are
\begin{eqnarray}
\dot{x}&=&\frac{\Delta l}{2\hbar}\sin{kl},\label{s22}\\
\dot{y}&=&\frac{\hbar}{m}k_y,\label{s23}\\
\dot{z}&=&\frac{\hbar}{m}k_z,\label{s24}\\
\hbar\dot{k}&=&eF-\frac{e\hbar k_y}{m}B\sin{\theta},\label{s25}\\
\hbar\dot{k_y}&=&-\left(\frac{\hbar k_z}{m}\cos{\theta}-\frac{\Delta l}{2\hbar}
\sin{kl}\sin{\theta} \right)eB,\label{s26}\\
\hbar\dot{k_z}&=&eF_z+\frac{eB}{m}\hbar k_y\cos{\theta}.Ê\label{s27}
\end{eqnarray}
Here $m$ is the effective mass of the electron, the magnetic field is $B(\cos\theta,0,\sin\theta)$ and, assuming that the electron density is independent of $y$ (see below), the electric field is $-(F,0,F_z)$. $\theta$ is the tilting angle between the magnetic field and the growth direction $x$. Note that \eqref{s26}, \eqref{s22} and \eqref{s24} produce the constant of motion $\hbar k_y+eB(z\cos\theta-x\sin\theta)$, so that
\begin{equation}
k_y= \frac{eB}{\hbar}(x\sin\theta-z\cos\theta)+K,\label{s28}
\end{equation}
where $K$ is a constant. We can set $K=0$ for appropriate initial conditions. Then the electron has only two degrees of freedom. Assuming $F_z=0$, we can combine \eqref{s27}, \eqref{s26}, and  \eqref{s25} to obtain 
\begin{eqnarray}
&&\ddot{k}_z+\left(\frac{eB}{m}\cos\theta\right)^2\! k_z=-\frac{e^2B^2\Delta l\sin 2\theta}{4m\hbar^2}\nonumber\\
&&\times\sin(k_zl\tan\theta-\omega_Bt-\phi),\label{s29}\\
&& \phi=k(0)l-k_z(0)l\tan\theta.\nonumber
\end{eqnarray}
For the tight-binding dispersion relation, this equation is the same as (2) in \cite{fro04}. However the assumption $F_z=0$ is incorrect. In fact, due to the single electron equations of motion \eqref{s22}-\eqref{s28}, the electron density should depend on $x$ and $z$, thereby producing a self-consistent electric potential that also depends on $x$ and $z$. Then $F_z\neq 0$.

\subsection{Boltzmann-Poisson system}
In the presence of a magnetic field, the BTE should have the single electron equations of motion as its characteristics. Then we will replace \eqref{s1} by \eqref{eq1}, in which $k_y$ is given by \eqref{s28} with $K=0$, and $\nu_e$ and $\nu_p$ are the phonon and impurity collision frequencies, respectively. We also have 
\begin{equation}
\mathcal{A}f= \frac{f(k)-f(-k)}{2},\label{s31}
\end{equation}
as we assume that the energy conserving impurity collisions only change momentum along the growth axis \cite{BGr05}. The relation between the 2D distribution function $f(x,z,k,k_z,t)$ and the 3D distribution function is 
\begin{eqnarray}
f^{3D}(x,y,z,k,k_y,k_z,t)= \frac{2\pi}{L_y}\, f(x,z,k,k_z,t)\nonumber\\
\times\delta\!\left(k_y-\frac{eB}{\hbar}(x\sin\theta-z\cos\theta)\right)\!, \label{s32}
\end{eqnarray}
in which $L_y$ is the large SL transversal length. The electric field is coupled to \eqref{s31} through the Poisson equation for the potential 
\begin{eqnarray}
&&\frac{\partial^2W}{\partial x^2}+\frac{\partial^2W}{\partial z^2}=\frac{e}{\varepsilon}(n-N_D),\label{s33}\\
\label{s35} && F=\frac{\partial W}{\partial x}, \quad F_z=\frac{\partial W}{\partial z},\\
&& n= \frac{2}{(2\pi)^2\!L_y}\!\int_{-\pi/l}^{\pi/l}\!\int\! f(x,z,k,k_z,t) dk dk_z.\label{s34}
\end{eqnarray}
Here the electron density has units of 1/(length)$^3$ and the factor 2 is due to spin degeneracy. The 2D distribution function $f$ is dimensionless. The 2D local equilibrium function $f^B$ should produce the same electron density as \eqref{s34}. From the 3D dispersion relation $(\Delta\cos kl)/2+\hbar^2(k_y^2+ k_z^2)/(2m)$ and integrating over $k_y$ a 3D local equilibrium similar to \eqref{s32}, we find \eqref{eq2}-\eqref{eq3}. 

\subsection{Esaki-Tsu stationary distribution function}
Let us set $\nu_p=0$ and repeat the derivation of \eqref{s11} for the BTE \eqref{eq1}. The equations of the characteristics are the single electron equations \eqref{s22}, \eqref{s24}, \eqref{s25} and \eqref{s27} and \eqref{s28} with $K=0$. Assume their solutions for fixed functions $F(x,z)$, $F_z(x,z)$ are
\begin{eqnarray}\nonumber
&&x=X(t-t_0;x_0,z_0,k_0,k_{z0}), \quad z=Z(t-t_0;x_0,z_0,k_0,k_{z0}),\\ 
&&k=K(t-t_0;x_0,z_0,k_0,k_{z0}),\quad k_z=K_z(t-t_0;x_0,z_0,k_0,k_{z0}),\nonumber \label{s38}
\end{eqnarray}
such that
\begin{eqnarray}\nonumber
X(0;x_0,z_0,k_0,k_{z0})=x_0, \, Z(0;x_0,z_0,k_0,k_{z0})=z_0,\\ K(0;x_0,z_0,k_0,k_{z0})=k_0,\, K_z(0;x_0,z_0,k_0,k_{z0})=k_{z0}.\label{s39}\nonumber
\end{eqnarray}
The distribution function is 
\begin{widetext}
\begin{eqnarray}\nonumber
f(t;x_0,z_0,k_0,k_{z0})&=&f_0(x_0,z_0,k_0,k_{z0}) e^{-(t-t_0)/\tau}\\
&+&\int_0^{(t-t_0)/\tau}\! e^{-\xi}f^B\!\left(K(t-t_0-\tau\xi;x_0,z_0,k_0,k_{z0}),K_z(t-t_0-\tau\xi;x_0,z_0,k_0,k_{z0})\right)\! d\xi.\label{s40}
\end{eqnarray}
\end{widetext}
To get the solution of the initial value problem for \eqref{eq1}, we have to solve first \eqref{s39} for $x_0$, $z_0$, $k_0$ and $k_{z0}$ as functions of $x$, $z$, $k$ and $k_z$:
\begin{eqnarray}
x_0=\mathcal{X}(x,z,k,k_z,t-t_0),\, z_0=\mathcal{Z}(x,z,k,k_z,t-t_0),\nonumber\\
 k_0=\mathcal{K}(x,z,k,k_z,t-t_0),\, k_{z0}=\mathcal{K}_z(x,z,k,k_z,t-t_0).\nonumber 
\end{eqnarray}
We now substitute these functions in the distribution function and set $t_0\to-\infty$. The result is the sought stationary distribution function provided this limit exists. See chapter 2 of \cite{hya09} for a similar study based on the {\em space independent} BTE under a {\em space independent} electric field. For $B=0$, we obtain the ETF from \eqref{s40}. However, it is clear that this procedure is rather cumbersome for $B\neq 0$ and $F_z\neq 0$. Moreover, using the resulting stationary distribution to get a drift-Poisson system needs justification.

\subsection{Leading order current density}
We now find an approximation to the solution of \eqref{eq1} that produces an approximate current density $(J_{nx},J_{nz})$. Firstly, it is convenient to define an auxiliary electromagnetic potential $\Omega=W-\hbar^2k^2_{y}/(2me)$ so that the Boltzmann-Poisson system becomes
\begin{widetext}
\begin{eqnarray}
&&\frac{\partial f}{\partial t} + \frac{\Delta l}{2 \hbar}\sin{kl}\frac{\partial f}{\partial x}+\frac{\hbar k_z}{m}
\frac{\partial f}{\partial z}+\frac{e}{\hbar}\frac{\partial\Omega}{\partial x}\frac{\partial f}{\partial k}+ \frac{e}{\hbar}\frac{\partial\Omega}{\partial z}\frac{\partial f}{\partial k_z} =\nu_ef^B-(\nu_e+
\nu_p\mathcal{A})f,\label{s42}\\
&&\varepsilon\left(\frac{\partial^2\Omega}{\partial x^2}+\frac{\partial^2\Omega}{\partial z^2}\right)\!=e\left(n-N_D-\frac{\varepsilon B^2}{m}\right)\!,\quad\Omega=W-\frac{\hbar^2k^2_{y}}{2me}=W-\frac{eB^2}{2m}(x\sin\theta-z\cos\theta)^2\! .\label{s43}
\end{eqnarray}\end{widetext}
The idea is that the terms containing the Lorentz force should balance the collision terms, the so-called hyperbolic limit \cite{BEP}:
\begin{eqnarray}
\frac{e}{\hbar}\frac{\partial\Omega}{\partial x}\frac{\partial f^{(0)}}{\partial k}\!+\! \frac{e}{\hbar}\frac{\partial\Omega}{\partial z}\frac{\partial f^{(0)}}{\partial k_z}\! =\!\nu_ef^B\nonumber\\
 - (\nu_e+\nu_p\!\mathcal{A})\, f^{(0)}.\label{s44}
\end{eqnarray}
We can solve these equations for a periodic function of $k$:
\begin{eqnarray}
 f^{(0)}(k,k_z)&=&\!\sum_{j=-\infty}^\infty e^{ijkl} f^{(0)}_j(k_z),\nonumber\\ 
 f^{(0)}_j(k_z)\!&=&\!\varphi_j(k_z)+i\psi_j(k_z)\nonumber\\
 &=&\! \frac{l}{2\pi}\int_{-\pi/l}^{\pi/l} f^{(0)}(k,k_z)e^{-ijkl}dk.\label{s45}
\end{eqnarray}
Inserting \eqref{s45} into \eqref{s44} and separating real and imaginary parts, we get 
\begin{widetext}
\begin{eqnarray}
&&\nu_e\varphi_j+\frac{e}{\hbar}\frac{\partial\Omega}{\partial z}\frac{\partial\varphi_j}{\partial k_z} - \frac{ej}{\hbar}\frac{\partial\Omega}{\partial x}\psi_j =\nu_e\hbar lL_y n \sqrt{\frac{\pi}{2mk_BT}}\frac{I_j\!\left(\frac{\Delta}{2k_BT}\right)\!}{I_0\!\left(\frac{\Delta}{2k_BT}\right)\!}  \exp\!\left(-\frac{\hbar^2k_z^2}{2mk_BT}\right)\!,\label{s46}\\
&&\frac{ejl}{\hbar}\frac{\partial\Omega}{\partial x}\varphi_j +(\nu_e+\nu_p)\psi_j+\frac{e}{\hbar}\frac{\partial\Omega}{\partial z}\,\frac{\partial\psi_j}{\partial k_z}=0.\label{s47}
\end{eqnarray}\end{widetext}
Taking the Fourier transform of these expressions and solving the resulting algebraic equations, we find 
\begin{widetext}
\begin{eqnarray}
&&\hat{\varphi}_j(\zeta)=n\frac{I_j\!\left(\frac{\Delta}{2k_BT}\right)\!}{I_0\!\left(\frac{\Delta}{2k_BT}\right)\!}\frac{\pi\nu_elL_y\!\left(\nu_e+\nu_p-\frac{ie\zeta}{\hbar}\frac{\partial\Omega}{\partial z} \right)\exp\!\left(-\frac{mk_BT\zeta^2}{2\hbar^2}\right)\!}{\nu_e(\nu_e+\nu_p)+\frac{e^2}{\hbar^2}\!\left[\!\left(jl\frac{\partial\Omega}{\partial x} \right)^2-\!\left(\zeta\frac{\partial\Omega}{\partial z}\right)^2\!\right]\!-(2\nu_e+\nu_p)\frac{ie\zeta}{\hbar}\frac{\partial\Omega}{\partial z} },\label{s48}\\
&&\hat{\psi}_j(\zeta)=-n\frac{I_j\!\left(\frac{\Delta}{2k_BT}\right)\!}{I_0\!\left(\frac{\Delta}{2k_BT}\right)\!}\frac{\pi\nu_e L_yjl^2\frac{e}{\hbar}\frac{\partial\Omega}{\partial x}\,\exp\!\left(-\frac{mk_BT\zeta^2}{2\hbar^2}\right)\!}{\nu_e(\nu_e+\nu_p)+\frac{e^2}{\hbar^2}\!\left[\!\left(jl\frac{\partial\Omega}{\partial x} \right)^2-\!\left(\zeta\frac{\partial\Omega}{\partial z}\right)^2\!\right]\!-(2\nu_e+\nu_p)\frac{ie\zeta}{\hbar}\frac{\partial\Omega}{\partial z} },\label{s49}
\end{eqnarray}\end{widetext}
in which 
\begin{eqnarray}
\hat{\varphi}_j(\zeta)=\int_{-\infty}^\infty e^{i\zeta k_z}\varphi_j(k_z)\, dk_z,\nonumber\\
\hat{\psi}_j(\zeta)=\int_{-\infty}^\infty e^{i\zeta k_z}\psi_j(k_z)\, dk_z.\label{s50}
\end{eqnarray}
The approximate current densities follow from \eqref{eq6}, \eqref{s48}, \eqref{s49} and \eqref{s50}:
\begin{eqnarray}
J_{nx}=-\frac{e\Delta\hat{\psi}_1(0)}{2\pi\hbar L_y} 
= env_p
\frac{2\frac{el\tau}{\hbar}\frac{\partial\Omega}{\partial x} }{1+\!\left(\frac{el\tau}{\hbar}\frac{\partial\Omega}{\partial x} \right)^2} , \label{s51}
\end{eqnarray}
\begin{eqnarray}
&&J_{nz}= -\frac{ie\hbar}{\pi mlL_y}\hat{\varphi}'_0(0)= \frac{e^2n}{m\nu_e}\frac{\partial\Omega}{\partial z},  \label{s52}
\end{eqnarray}
\begin{eqnarray} \label{s53}
v_p=\frac{\Delta l}{4\hbar}\frac{I_1\!\left(\frac{\Delta}{2k_BT}\right)\!}{I_0\!\left(\frac{\Delta}{2k_BT}\right)\!}\, \sqrt{ {\frac{\nu_e}{\nu_e+\nu_p} } }, \end{eqnarray} 
\begin{eqnarray}
\tau=\frac{1}{\sqrt{\nu_e(\nu_e+\nu_p)}}.  \label{s54}
\end{eqnarray} 
For $B=0$, \eqref{s51} yields the temperature dependent ETDV \eqref{s16} with peak velocity \eqref{s53}. The latter equation generalizes \eqref{s30} to the case of phonon and impurity collisions, and it becomes the latter for $\nu_p=0$. For $B=0$, \eqref{s52} yields $J_{nz}=0$. Note that replacing $\tau$ by \eqref{s54} in the generalized ETF \eqref{s12}, and multiplying it by $\delta=1/\sqrt{1+\nu_p/\nu_e}$, we obtain \eqref{s53}; cf \cite{fro04}.

\end{document}